\documentclass[12pt]{article}
\usepackage{epsf,amsfonts,latexsym}
\newsymbol\ltimes 226E
\newsymbol\rtimes 226F
\epsfverbosetrue \textheight=22.7cm \textwidth=16.cm
\def\goth{\mathfrak}
\def\double{\mathbb}
\def\ccal{\cal}

\def\cc{{\double C}}

\def\rr{{\double R}}
\def\zz{{\double Z}}

\def\aa{{\cal A}}

\def\dd{{\cal D}}
\def\gg{{\goth g}}
\def\hh{{\cal H}}

\def\ff{{\cal F}}
\def\mm{{{\ccal M}}}

\def\aa{{\cal A}}
\def\dd{{\cal D}}
\def\hh{{\cal H}}
\def\ff{{\cal F}}

\def\sss{{\cal S}}

\def\t{{\rm tr}\,}

\def\ddd{{\,\hbox{$\partial\!\!\!/$}}}

\def\dee{\hbox{\rm D}}
\def\de{\hbox{\rm d}}

\def\pa{{\partial}}

\def\lb{\left[}
\def\rb{\right]}

\def\ker{{\rm ker}}
\def\ul{\underline}
\def\ot{\otimes}
\def\op{\oplus}

\def\bb{\begin{eqnarray}}
\def\ee{\end{eqnarray}}
\def\eee{\nonumber\end{eqnarray}}
\def\pp{\pmatrix}
\def\qq{\quad}

\newcommand{\CG}{\hbox{{$\cal G$}}} 

\newcommand{\CA}{\hbox{{$\cal A$}}}
\newcommand{\CC}{\hbox{{$\cal C$}}}

\newcommand{\R}{\mathbb{R}}
\newcommand{\C}{\mathbb{C}}
\newcommand{\Z}{\mathbb{Z}}
\newcommand{\quat}{\mathbb{H}}

\def\del{\partial}
\newcommand{\h}{{\scriptstyle\frac{1}{2}}}
\newcommand{\extd}{{\rm d}}

\newcommand{\isom}{{\cong}}

\newcommand{\tens}{\mathop{\otimes}}

\newcommand{\id}{{\rm id}}
\newcommand{\<}{\langle}
\renewcommand{\>}{\rangle}

\newcommand{\proof}{{\bf Proof\ }}
\newcommand{\eproof}{$\quad \diamond$}

\newcommand{\eqn}[2]{\begin{equation}#2\label{#1}\end{equation}}

\newcommand{\alignn}[1]{\begin{eqnarray*}#1\end{eqnarray*}}

\newtheorem{lemma}{Lemma}[section]
\newtheorem{propos}[lemma]{Proposition}

\begin{document}

\hsize 17truecm \vsize 24truecm \font\twelve=cmbx10 at 13pt
\font\eightrm=cmr8 \baselineskip 18.3pt

{\ }\qquad \hskip 4.3in \vspace{.2in}

\begin{center} {\LARGE $\Z_2\times\Z_2$ Lattice as a
Connes-Lott-Quantum Group Model} \\
\baselineskip 13pt{\ }\\ {\ }\\ Shahn Majid \\ {\ }\\ School of
Mathematical Sciences\\ Queen Mary, University of London\\Mile
End Rd, London E1 4NS, UK
\\{\ }\\ +\\{\ }\\ Thomas Sch\"ucker\\ {\ }\\Centre de Physique
Th\'eorique\\ case 907\\ F-13288 Marseille, cedex 9
\end{center}
\begin{center} December 2000
\end{center}

\begin{quote}\baselineskip 13pt
\noindent{\bf Abstract} We apply quantum group methods for
noncommutative geometry to the $\Z_2\times\Z_2$ lattice to obtain
a natural Dirac operator on this discrete space. This then leads
to an interpretation of the Higgs fields as the discrete part of
spacetime in the Connes-Lott formalism for elementary particle
Lagrangians. The model provides a setting where both the quantum
groups and the Connes approach to noncommutative geometry can be
usefully combined, with some of Connes' axioms, notably the
first-order condition, replaced by algebraic methods based on the
group structure. The noncommutative geometry has nontrivial
cohomology and moduli of flat connections, both of which we
compute.
\end{quote}

\section{Introduction}

In \cite{cl}\cite{co} Connes and Lott proposed a framework for the
standard model in elementary particle physics based on a discrete
and typically noncommutative part adjoined to conventional
spacetime. Fields on this composite spacetime appear as
multiplets of fields on ordinary spacetime and, for the right
choice of discrete part, one obtains exactly the standard model
of particle physics. The Dirac operator on the discrete part
encodes the masses of fermions on usual spacetime. This approach
`packages' the standard model into an elegant framework where
also the Higgs field arises naturally. However, most of the
parameters of the standard model are still left undetermined
because in the Connes approach to noncommutative geometry almost
any self-adjoint operator $\ddd$ can be taken on the discrete
part of spacetime in the role of Dirac.

Meanwhile coming from quantum groups is a `constructive' approach
to noncommutative geometry which includes also finite groups and
other discrete spaces. In this approach, because of the existence
of q-deformed examples one keeps `eye contact' with conventional
geometric ideas and thereby builds up the different layers of
(noncommutative) geometry up to and including, in recent
work\cite{Ma:rieq}, the Dirac operator. In other words when the
Connes-Lott formalism and the quantum groups formalism are
combined one has natural `geometric' criteria for the choice of
Dirac operator on the discrete spacetime which translates
directly into predictions in elementary particle physics.

In this paper we develop a nontrivial model for which these two
approaches can be combined in this way, and explore fully both
approaches for this model. The model has `discrete part'
$\Z_2\times\Z_2$ which has a commutative coordinate algebra but
which we equip with noncommutative differentials coming naturally
from the quantum groups approach (a bicovariant differential
calculus in the sense of \cite{Wor:dif}). The model is too simple
to lead to exactly the standard model (for this one wants the
noncommutative algebra $\C\oplus\quat\oplus M_3$) but it exhibits
many of the same features. Moreover, the model is of independent
interest as a discrete (lattice) model of spacetime useful in a
variety of other contexts, e.g. potentially for QCD.

In Section~2 we explore the model using Connes
formalism\cite{cl}\cite{co}. Thus, starting with the bicovariant
differential calculus suggested by quantum group methods, we take
the natural 2-dimensional Dirac operator and apply the method of
Connes to induce an entire exterior algebra, Hodge $*$ and other
constructions on this discrete 2-d `spacetime' (not to be
confused with conventional spacetime of course but thought of in
that way). We find a Higgs-effect and aspects of symmetry
breaking on this discrete spacetime. Following Connes, we work
very explicitly with 1-forms and 2-forms etc as certain concrete
matrices. The higher forms are not so easily computed by these
methods, however.

In Section~3 we construct this exterior algebra etc induced by
$\ddd$ more algebraically point using quantum group methods. Here
the exterior algebra is obtained as a quotient of the universal
differential calculus by generators and relations, and not
concretely as certain matrices. We show how many of the
computations in the Connes-Lott model building kit can be done
more in line with classical constructions using these algebraic
quantum group methods. Using these methods we are then able to
take the computations of Section~2 much further. We fully compute
the exterior algebra, its quantum de Rham cohomology and its
moduli of zero curvature gauge fields, all of which turn out to be
nontrivial. We note that quantum group methods for the
noncommutative geometry on finite groups have recently been
developed in some generality\cite{Ma:rieq}\cite{Ma:cons},
including  gravity and a first contact with Connes' method which
we use now (an analysis of the 2-forms). See also
\cite{MaRai:ele} where the cohomology and gauge theory for the
permutation group $S_3$ is recently computed. The
$\Z_2\times\Z_2$ model can be viewed as another nontrivial
noncommutative geometry in this family. The use of noncommutative
geometry for discrete spacetimes itself originates in the bilocal
nature of finite difference differentials, and is quite
fundamental.

In Section~4 we return to the physics by combining this discrete
2d spacetime with conventional spacetime to pull out the
resulting fairly straightforward model of particle physics and
some predictions ensuing form our particular `geometrical' choice
of $\ddd$. In Section~5 we look at a further chapter of Connes
method\cite{dix}\cite{cc}, namely the spectral action and gravity,
automorphisms and spin automorphisms. The $\Z_2\times \Z_2$ model
is again simple enough that all aspects are computable explicitly.

Finally, in Section~6 we conclude with some comments on the
general lattice $(\Z_m)^n$. Using again our algebraic quantum
group methods we show that similar features hold for higher
dimensional $(\Z_2)^n$ but that on the other hand for $m>2$ the
nontrivial features of the model such as
 the Higgs potential disappear, i.e. are a very specific
to the use of $\Z_2$.

 \section{The $2\times 2$ lattice \`a la Connes-Lott}

In this section we apply the Connes-Lott model building kit
\cite{cl} to a $2\times 2$ lattice described by the associative,
unital star algebra \bb\aa=\cc[\zz_2\times\zz_2]\ \owns\
f(x,y),\qq x,y=0,1\ {\rm mod}\,2.\ee We define right translations
in $x$ and $y$ direction by \bb
(R_xf)(x,y):=f(x+1,y),\qq(R_yf)(x,y):=f(x,y+1),\ee and the partial
derivatives by \bb \pa_x:=R_x-1,\qq\pa_y:=R_y-1.\ee The following
relations will be useful,
\bb (R_x)^2=1,&(R_y)^2=1,&R_xR_y=R_yR_x,\\
(\pa_x)^2=-2\pa_x,&(\pa_y)^2=-2\pa_y,& \pa_x\pa_y=\pa_y\pa_x,\ee
and the Leibniz rule \bb\pa_x(fg)=(\pa_xf)g+(R_xf)\pa_xg=
(\pa_xf)R_xg+f\pa_xg.\ee We define the Hilbert space of spinors,
\bb\hh:=\aa\ot\cc^2\ \owns\psi=\pp{\psi_R\cr
\psi_L},\qq\psi_R,\psi_L \in\aa,\ee with scalar product \bb
(\psi,\tilde\psi):=\sum_{x,y=0}^1 [\bar\psi_R(x,y)\tilde
\psi_R(x,y)+ \bar\psi_L(x,y)\tilde \psi_L(x,y)]\ee and the
faithful representation $\rho$ of the algebra $\aa$ on $\hh$ by
pointwise multiplication, \bb
(\rho(f)\psi)(x,y):=f(x,y)\psi(x,y).\ee We will need the relation
\bb (R_x\ot1_2)\rho(f)=\rho(R_xf)(R_x\ot1_2).\ee The third input
item is the Dirac operator that we take to be the lattice Dirac
operator, \bb \ddd:=\pa_x\ot\gamma^x+\pa_y\ot\gamma^y\ee with the
Hermitian Pauli matrices \bb\gamma^x:=\pp{0&1\cr 1&0},\qq
\gamma^y:=\pp{0&-i\cr i&0},\qq \gamma^3:=\pp{1&0\cr 0&-1}.\ee
They satisfy \bb (\gamma^x)^2=(\gamma^y)^2=1_2,\qq
\gamma^x\gamma^y=-\gamma^y\gamma^x= i\gamma^3.\ee Note that this
Dirac operator is self--adjoint without an imaginary $i$ in
front. Note also that this Dirac operator like any lattice Dirac
operator cannot satisfy Connes' \cite{co} first order condition
\cite{gs}. Nevertheless we can use these three items, the algebra
$\aa$, its representation on $\hh$ and the Dirac operator $\ddd$
to construct a Connes-Lott model \cite{cl}.

 The first step involves an auxiliary differential algebra
$\Omega_{univ}(\aa),$ the universal exterior algebra of $\aa$: \bb
\Omega^0_{univ}(\aa) := \aa,\ee $\Omega^1_{univ}(\aa)$ is spanned
over $\aa$ by symbols $\extd a$, $ a \in \aa$ with relations \bb
\extd 1 = 0,\qq \extd(ab) = (\extd a)b+a\extd b.\ee Therefore it
consists of finite sums of terms of the form
 $a_0\extd a_1,$
\bb   \Omega^1_{univ}(\aa) = \left\{ \sum_j a^j_0\extd
a^j_1,\quad a^j_0, a^j_1\in \aa\right\}\ee and likewise for
higher $p$, \bb \Omega^p_{univ}(\aa)= \left\{ \sum_j a^j_0\extd
a^j_1...\extd a^j_p ,\quad a^j_q\in \aa\right\}.\ee
 The differential $\extd$ is defined by
$   \extd(a_0\extd a_1...\extd a_p) :=
   \extd a_0\extd a_1...\extd a_p.$ The involution
$^*$ is extended from the algebra $\aa$ to
 $\Omega^1_{univ}(\aa)$ by putting
\bb   (\extd a)^* := -\extd(a^*). \ee Note that it can also be
useful to define $(\extd a)^* := \extd(a^*)$ which amounts to
replacing $\extd$ by $-i\extd.$ With the definition $
(\alpha\beta)^*=\beta^*\alpha^*$ for forms $\alpha,\beta$, the
involution is extended to the whole universal exterior algebra.

The next step is to extend the representation $\rho$ on $\hh$
from the algebra $\aa$   to its universal exterior algebra. This
extension is the central piece of Connes' algorithm: \bb && \pi :
\Omega_{univ}(\aa)  \longrightarrow{\rm End}(\hh)
   \cr
 &&\pi(a_0\extd a_1...\extd a_p) :=\rho(a_0)[\ddd,\rho(a_1)]
...[\ddd,\rho(a_p)] \label{pi}.\ee A straightforward calculation
shows that $\pi$ is in fact a representation of
$\Omega_{univ}(\aa)$ as an algebra with involution, and we are
tempted to define also a differential, denoted again by $\de $,
 on the images $\pi(\Omega^p_{univ}(\aa))$ in each degree by
\bb \de \pi(\alpha):=\pi(\extd\alpha),\quad \forall
\alpha\in\Omega^p_{univ}(\aa). \label{trial}\ee However, this
definition does not make sense if there are forms
$\alpha\in\Omega_{univ}(\aa)$ with $\pi(\alpha)=0$ and
$\pi(\extd\alpha) \not= 0$. By dividing out these unpleasant
forms, Connes constructs a new differential algebra
$\Omega_\ddd(\aa)$, the interesting object: \bb \Omega_\ddd(\aa)
:= {{\pi\left(\Omega_{univ}(\aa)\right)}\over {\cal J}}\ee with
\bb {\cal J} := \pi\left(\extd\ker\pi\right) =: \bigoplus_p {\cal
J}_p\ee (${\cal J}$ for junk). On the quotient now, the
differential (\ref{trial}) is well defined. Degree by degree we
have:  \bb \Omega_\ddd^0(\aa) =\rho(\aa)\ee    because ${\cal
J}^0=0$ , \bb \Omega_\ddd^1(\aa) = \pi(\Omega^1_{univ}(\aa))\ee
because $\rho$ is faithful, and in degree $p\geq2$, \bb
\Omega_\ddd^p(\aa) = {{\pi(\Omega^p_{univ}(\aa))}\over
{\pi(\extd\ker\pi_{p-1})}}.\ee Here $\pi_{p-1}$ denotes $\pi$
restricted to degree $p-1$ forms. We remind the motivation of
Connes' construction: in the continuum, $\aa$ the algebra of
differentiable functions on the 2-torus and $\ddd$ the genuine
Dirac operator, $\Omega_\ddd(\aa)$ is de Rham's exterior algebra
of differential forms.

In our lattice model all forms are explicit $8\times 8$ matrices.
E.g. in equation (\ref{omegax}) below \bb \omega _x=\pp{0&1&0&0\cr
1&0&0&0\cr 0&0&0&1\cr 0&0&1&0}\ot\pp{0&1\cr 1&0}\ee with respect
to the basis $\{\delta _{00},\delta _{10},\delta _{01}, \delta
_{11}\}\ot\left\{ \pp{1\cr 0},\pp{0\cr 1}\right\} $. Let us
compute the 1-forms, \bb\pi(\extd f)=
[\ddd,\rho(f)]=\rho(\pa_xf)R_x\ot\gamma^x
+\rho(\pa_yf)R_y\ot\gamma^y.\ee We denote by
$\delta_{00},\delta_{10},\delta_{01},\delta_{11}\in\aa$ the four
delta functions, $\delta_{00}$ for instance is one on $x=0,y=0$
and zero on the other three points. Then $\Omega^1_\ddd(\aa)$ is
spanned over $\aa$ by the two elements, \bb
\omega_x:=\pi(\pa_xX\extd X)=R_x\ot\gamma^x,
\qq\omega_y:=\pi(\pa_yY\extd Y)=R_y\ot\gamma^y \label{omegax}\ee
where we have put $X:=\delta_{10}+\delta_{11}$ and $Y:=
\delta_{01}+\delta_{11}$. Our generators are Hermitian,
$\omega_x^*=\omega_x,\ \omega_y^*=\omega_y$.

The 2-forms are represented by \bb\pi(\extd f\extd g)=
\rho(\pa_xfR_x\pa_xg+\pa_yfR_y\pa_yg)\ot1_2+
\rho(\pa_xfR_x\pa_yg-\pa_yfR_y\pa_xg)R_xR_y\ot
\gamma^x\gamma^y\ee and a straight forward calculation in a basis
using $\delta$ functions shows that in degree two the junk
vanishes, ${\cal J}_2=0$. Therefore we get \bb\label{matrels}
(\omega_x)^2=(\omega_y)^2=1,\qq
\omega_x\omega_y=-\omega_y\omega_x, \qq\de\omega_x=\de\omega_y=2
.\ee

At this stage there is a first contact with gauge theories.
Consider the vector space of Hermitian 1-forms $ \left\{ H\in
\Omega_\ddd^1(\aa),\ H^*=H \right\}.$ A general  element $H$ is
of the form \bb   H = \rho(h_x)\omega_x+\rho(h_y)\omega_y,\ee
with \bb h_x(0,0)=h_x(1,0)^*,\qq h_x(1,1)=h_x(0,1)^*,\qq
h_y(0,0)=h_y(0,1)^*,\qq h_y(1,1)=h_y(1,0)^*.\qq\ee These elements
$H$ are gauge potentials on the lattice. In fact the space of
gauge potentials carries an affine representation of the group of
unitaries \bb    U(\aa): = \{u\in \aa,\ uu^* =u^*u=1\}\ =:\CG.\ee
In our example this group is Maxwell's local $U(1)$. In general
its action is defined by \bb H^u &:=&\
\rho(u)H\rho(u^{-1})+\rho(u)\de(\rho(u^{-1})) \cr
       &=&\
\rho(u)H\rho(u^{-1})+\rho(u)[\ddd,\rho (u^{-1})] \cr
      &=&\ \rho(u)[H+\ddd]\rho(u^{-1})-\ddd.\ee
$H^u$ is the ``gauge transformed of $H$''.  As usual every gauge
potential $H$ defines a covariant derivative  $\de+H$, covariant
under the left action of $\CG$ on $\Omega_\ddd\aa$: \bb
^u\omega := \rho(u)\omega, \quad \omega\in\Omega_\ddd\aa\ee
 which means
\bb   (\de +H^u)\ ^u\omega = \ ^u\lb(\de +H)\omega\rb.\ee
 Also we define the curvature $C$ of $H$ by
\bb   C := \de H+H^2\ \in\Omega_\ddd^2(\aa).\ee The curvature $C$
is a Hermitian 2-form with {\it homogeneous} gauge transformations
 \bb   C^u :=
\de (H^u)+(H^u)^2 = \rho(u) C \rho(u^{-1}).\ee In our example we
get \bb C&=& \rho(\pa_xh_x+\pa_yh_y+2h_x+2h_y
+h_xR_xh_x+h_yR_yh_y)\,\omega_x^2\cr
&&+\rho(-\pa_yh_x+\pa_xh_y+h_xR_xh_y-h_yR_yh_x)\,
\omega_x\omega_y.\ee In the last step we construct the Yang-Mills
action. To this end we need a scalar product on the space of
2-forms. But our forms are operators on the finite dimensional
Hilbert space $\hh$ and we have a natural scalar product.

At this point we must note that although $\rho $ is faithful $\pi
$ is not, not even after dividing out the junk. And even worse,
the image in End$(\hh)$ does not remember its degree. (This
complication does not occur in the continuum.) In our example for
instance we meet the $8\times 8$ unit matrix as 0-form $\rho (1)$
and as 2-form $\omega _x^2$. By definition the scalar product of
two forms of different degree is taken to be zero, for forms of
same degree $p$, we define \bb (\omega ,\tilde \omega ):= \t
(\omega ^*\tilde \omega ),\  \omega ,\tilde \omega \in
\Omega^p_\ddd(\aa).\ee For example, $\rho (1),\ \omega_x,\
\omega_y,\ \omega_x \omega_y,\ \omega_x^2 $ are orthogonal
generators, all normed to $\sqrt 8$. More generally we have \bb
(\rho (f),\rho (\tilde f))=2(f,\tilde f),\qq (\rho
(f)\omega_x,\rho (\tilde f)\omega_x)=2(f,\tilde f),\qq (\rho
(f)\omega_y,\rho (\tilde f)\omega_y)=2(f,\tilde f),\qq\ee\bb
(\rho (f)\omega_x\omega_y,\rho (\tilde
f)\omega_x\omega_y)=2(f,\tilde f),\qq (\rho (f)\omega_x^2,\rho
(\tilde f)\omega_x^2)=2(f,\tilde f),\qq\ee with \bb (f,\tilde
f):=\sum_{x,y=0}^1\bar f(x,y) \tilde f(x,y).\ee We are now in
position to define the Yang-Mills action $V_0(H)=(C,C)$. By
construction it is a positive, gauge invariant polynomial of
fourth order in the values of $h_x$ and $h_y$. Its minimum,
$H=0$, breaks the gauge invariance. In order to compute the Higgs
potential we introduce a new variable \cite{vs}, \bb
\varphi:=H+\ddd_{\CG},\ee with \bb
\ddd_{\CG}:=-\int_{\CG}\pi(u^{-1}\extd u)\de
u=\ddd-\int_{\CG}\rho(u^{-1})\ddd\rho(u)\de u=\omega_x+\omega_y\ee
and $\de u$ the normalised Haar measure of the compact Lie group
$\CG$. We decide that the Dirac operator does not transform under
gauge transformations. Then $\varphi$ transforms homogeneously,
\bb\varphi^u=\rho(u)\varphi\rho(u^{-1}).\label{hom}\ee
 Let us expand the homogeneous variable as
\bb \varphi=\rho(\varphi_x)\omega_x+\rho(\varphi_y)\omega_y,\ee
with $\varphi_x=h_x+1,\ \varphi_y=h_y+1.$ Then we can rewrite the
curvature as \bb C=\rho (\varphi_xR_x\varphi_x
+\varphi_yR_y\varphi _y-2)\,\omega_x^2 +\rho
(\varphi_xR_x\varphi_y-\varphi
_yR_y\varphi_x)\,\omega_x\omega_y,\ee and the Yang-Mills action
can be written explicitly: \bb V_0&=2\{&\ \ [|\varphi_x(0,0)|^2
+|\varphi_y(0,0)|^2-2]^2 +[|\varphi_x(0,0)|^2
+|\varphi_y(1,1)|^2-2]^2\cr &&+[|\varphi_x(1,1)|^2
+|\varphi_y(0,0)|^2-2]^2 +[|\varphi_x(1,1)|^2
+|\varphi_y(1,1)|^2-2]^2\cr && +|\varphi_x(0,0)\varphi_y(1,1)^*-
\varphi_x(1,1)^*\varphi_y(0,0)|^2 \cr && +|\varphi_x(0,0)\varphi
_y(0,0)^*- \varphi_x(1,1)^*\varphi_y(1,1)|^2\qq\}.
\label{brother}\ee The little group of its minimum $H=0$ or
$\varphi:=\ddd_{\CG}$ is the group of rigid $U(1)$
transformations as in the continuous case. However unlike in the
continuous case, there is a gauge invariant point, $\varphi=0$ or
$H=-\ddd_{\CG}$ which is also a local maximum of the Yang-Mills
action $V_0$. The existence of this gauge invariant point
indicates that in this model the gauge potential $H$
simultaneously plays the role of a Higgs scalar and the lattice
Yang-Mills action is its Higgs potential.

The minima of the potential $V_0$ are continuously degenerate,
$\varphi_x(0,0)=\varphi_x(1,1)=\sqrt 2 \sin\beta $, $\varphi
_y(0,0)=\varphi_y(1,1)=\sqrt 2\cos\beta $ All minima have little
groups is $U(1)$ except when $\beta$ is an integer multiple of
$\pi/2 $. Then the little group is $U(1)^2$. Let us remark that
this model is similar to example 3.1 in \cite{vs}: its algebra is
represented vectorially, but does not commute with the Dirac
operator and its potential has degenerate minima with different
little groups.

\section{Quantum group methods for the same model}

In the previous section we have pulled the partial derivatives
and Dirac operator `out of a hat' (motivated of course by the
wish to include lattice differentials). In particular, since the
resulting $\ddd$ does not obey the first order condition in
Connes's axioms in any standard way, it is not motivated from
that theory. Rather this choice of differentials comes from
requiring translation invariance under the group structure of
$G=\Z_2\times \Z_2$. This is part of the `quantum groups
approach' where one builds up the different layers of
noncommutative geometry based on the group or quantum group
structure. This approach also has a more algebraic way of working
in which we deal algebraically with the differential forms rather
than concretely as matrices. In this section we explain the
construction of the Dirac operator from the quantum groups point
of view and significantly extend the results of the previous
section using this algebraic language. In particular, it allows
us to compute the full exterior algebra and its cohomolgy as well
as the full moduli of flat connections in the gauge theory
picture that underlies the model. Note that our results here
should not be confused with the question of existence or not of a
spectral triple for a given differential calculus on a finite
group, e.g. as in \cite{PasSit} and elsewhere (we are interested
in a particular lattice Dirac operator $\ddd$).

We will use more algebraic notation. Thus, we work with the
universal exterior algebra $\Omega_{univ}(\aa)$ explicitly as an
algebra with a finite number of left-invariant 1-forms as
generators. We then exhibit $\Omega_\ddd(\aa)$ not as matrices
but as a quotient $\Omega(\aa)$ of the universal one by relations
among the generators (keeping the same names for the generators
in the quotient). For ease of reference, the resulting dictionary
with the concrete matrices in the previous section will be \bb
\omega_x=\pi(e_x), \quad \omega_y=\pi(e_y),\quad H=\pi(\alpha),
\quad C=\pi(F),\quad \varphi=\pi(\Phi),\ee for abstract forms
$e_x,e_y,\alpha,\Phi$ and $F$ in $\Omega(A)$.

\subsection{Exterior algebra and cohomology}

Thus, a differential calculus from the quantum groups point of
view means any $\CA-\CA$ bimodule $\Omega^1(\CA)$ and a map
$\extd:\CA\to\Omega^1(\CA)$ obeying the Leibniz rule. When $\CA$
is a Hopf algebra we demand further that $\Omega^1(\CA)$ is
bicovariant\cite{Wor:dif}. Just as a topological space can admit
more than one differential structure, one has to classify the
possible $\Omega^1(\CA)$. From results in \cite{Wor:dif} it is
immediate for the case of $\CA=\C[G]$ the functions on a finite
group, that the possible bicovariant calculi are in
correspondence with subsets \eqn{CCG}{ \CC\subset G,\quad
e\notin\CC} where $e$ is the group identity. The elements of
$\CC$ label the `basic 1-forms' $\{ e_a\}$ of the corresponding
$\Omega^1_{\CC}(\CA)$ and any other 1-form is a unique linear
combination of these with coefficients from $\CA$. The
commutation rules and general form of $\extd$ in this
construction are \eqn{dG}{  e_af=R_a(f) e_a,\quad \extd f
=\sum_{a\in \CC} (\del_a f) e_a,\quad \del_a=R_a-\id.} One of the
nice features of this construction is that it does not require
the group to be Abelian, i.e. extends to nonAbelian or `curved'
lattices.

Also, these calculi are all quotients of the universal
$\Omega^1_{univ}(\CA)$ which can either be defined `symbolically'
as in the previous section or very explicitly as the elements of
$\CA\tens \CA$ whose product is zero. Here $\extd f=1\tens
f-f\tens 1$. For functions on a finite set $G$ we take for $\CA$
a basis of $\delta$-functions and hence \bb
\Omega^1_{univ}(\CA)=\{\delta_g\tens\delta_h
=\delta_g\extd\delta_h|\
g\ne h,\ g,h\in G\}.\ee The quotient to our chosen
$\Omega^1_{\CC}(\CA)$ means to set to zero all such elements
except for those for which $(g,h)\in E$ some subset of allowed
directions. In the group case this subset $E$ is defined in a
translation-invariant manner from $\CC$, namely as pairs $(g,h)$
for which the difference (in the additive case) likes in $\CC$.

In our case we choose the subset \eqn{CC}{ \CC=\{x,y\},\quad
x=(1,0),\quad y=(0,1)} so every 1-form is uniquely of the form
\eqn{alpha}{ \alpha=\alpha_x e_x+\alpha_y e_y,\quad
\alpha_x,\alpha_y\in \CA.} The basic 1-forms can be written
explicitly as \eqn{basic}{  e_x=\sum_{g\in
\Z_2\times\Z_2}\delta_g\extd \delta_{g+x},\quad
e_y=\sum_{g\in\Z_2\times\Z_2}\delta_g\extd\delta_{g+y}.} This is
a full description of $\Omega^1_{\CC}(\CA)$ as defined by the
choice of $\CC$ above. Clearly we have the same answer as in
Section~2 where we posited an operator $\ddd$ and derived
$\Omega^1_\ddd(\CA)$, i.e. \eqn{isomd}{
\pi:\Omega^1_{\CC}(\CA)\isom\Omega^1_\ddd(\CA).} Actually this is
a well-known general feature; for any linearly independent
$\{\gamma^a\}$ and $\ddd=\del_a\gamma^a$ we will have the same
agreement between the Connes and the quantum groups approach up
to degree 1, by construction. We will work with this
$\Omega^1_{\CC}(\CA)$ and no longer
 write the $\CC$ explicitly.

Next we consider higher degree forms. For any first order calculus
$\Omega^1(\CA)$ there is a `linear prolongation' where we impose
only the relations in higher forms inherited from those at degree
1 and $\extd^2=0$. The latter in our case means \alignn{
0&&=\extd( \del_x (f) e_x+\del_y(f) e_y)\\ &&=-2\del_x(f)
e_x^2-2\del_y(f) e_y +\del_x\del_y(f)( e_x e_y+ e_y e_x)
+\del_x(f)\extd e_x+\del_y(f)\extd e_x} and choosing
$f=\delta_{00}+\delta_{01}-\delta_{10}-\delta_{11}$ which obeys
$\del_x f=-2f$ and $\del_yf=0$, and a similar function for the
roles of $x,y$ interchanged, one finds \eqn{maxrel}{\extd  e_x=2
e_x^2,\quad \extd e_y=2 e_y^2,\quad e_x e_y=- e_y e_x.} The last
of these follows from putting the first two into the $\extd^2=0$
equation and then choosing a function with $\del_x\del_yf\ne 0$.
Beyond this linear prolongation exterior algebra, we are free in
the `constructive' approach to impose further relations in higher
degrees. One general construction exists due to
Woronowicz\cite{Wor:dif} and for an Abelian group as in our case
it would simply imply that $e_x^2= e_y^2=0$. We do {\em not} do
this but instead impose the relation
 coming out of the Connes machinery in Section~2, namely
\eqn{conrelns}{ e_x^2=e_y^2} in the exterior algebra. Here the
Connes approach and the Woronowicz approach for higher
differentials diverge and we choose the former. Then
$\Omega^2(\CA)$ is 2-dimensional over $\CA$, being spanned by
$e_x e_y,e_y^2$. Choosing representatives in the universal
exterior algebra for these, our explicit calculations
(\ref{matrels}) in the previous section show that their images
under $\pi$ are linearly independent, hence \eqn{isomd2}{
\pi:\Omega^2(\CA)\isom\Omega^2_\ddd(\CA)} when constructed in
this way.

Next we take the `quadratic prolongation' of this
$\Omega^1,\Omega^2$ to degree 3 and higher, i.e. impose no
further relations than the quadratic ones (\ref{conrelns}) and
(\ref{maxrel}) already imposed and whatever is implied by these.

\begin{propos} The quadratic exterior algebra $\Omega(\CA)$
generated by $e_x,e_y$ with relations $e_x^2=e_y^2$ and
$\{e_x,e_y\}=0$ is isomorphic to $\Omega_\ddd(\CA)$. Moreover,
there is a generating 1-form
\[ \theta=e_x+e_y,\quad \extd
\alpha=\{\theta,\alpha]\] for all forms $\alpha$, where we use
commutator on even degree and anticommutator on odd degree and
$\pi(\theta)=\ddd_{\CG}$ as a matrix.
\end{propos}
\proof First we compute what this quadratic exterior algebra
looks like. We then compare it with Connes construction and check
the isomorphism. The remark about $\theta$ is then an immediate
corollary since it is a general feature of the linear
prolongation of $\Omega^1(\CA)$ (where we have seen that
$e_x,e_y$ anticommute) and hence holds in the quadratic  exterior
algebra quotient (as well as in the Woronowicz exterior algebra
where it is would be well-known). To compute the quadratic
exterior algebra we note that
\[ \extd (e_xe_y+e_ye_x)=2e_x^2e_y-e_x 2e_y^2+2e_y^2e_x-e_y
2e_x^2=0,\quad \extd(e_x^2-e_y^2)=0\] automatically, hence there
are no implied relations in degree 3 or higher coming from these.
In that case we have only the relations (\ref{conrelns}) and the
anticommutativity relations. From this it is easy to see that \bb
\Omega^p(\CA)=\CA\< e_xe_y^{p-1},\ e_y^p\>,\quad p\ge 1\ee is
2-dimensional over $\CA=\C[\Z_2\times\Z_2]$. By comparison we
recall Connes definition
\[ \Omega_\ddd^p(\CA)=\pi_p(\Omega_{univ}^p(\CA))/{\cal J}_p,\quad
{\cal J}_p=\pi_p(\extd \ker\pi_{p-1})\] where $\pi_p$ denotes
$\pi$ in degree $p$ of the universal calculus. The quadratic
exterior algebra is at least as big as the Connes one since it
uses only the relations already holding in the latter in degrees
1,2. Hence all that we really need to establish an isomorphism is
to show that $\Omega_{\ddd}^p$ has dimension 2 over
$\C[\Z_2\times Z_2]$. In fact it suffices to exhibit two elements
of the universal exterior algebra with linear independent images
in $\pi_p(\Omega_{univ}^p(\CA))$ for each $p$, after which the
result can be proven by induction. Indeed, knowing our result for
$\Omega_\ddd^{p-1}(\CA)\isom\Omega^{p-1}(\CA)$, we know that the
kernel of $\pi_{p-1}$ is generated by the quadratic relations
above. But $\extd$ of these, by the computation above, lies again
in the ideal generated by these relations, so ${\cal J}_{p}=0$
and hence $\Omega_\ddd^{p}(\CA)=\pi_{p}(\Omega_{univ}^{p}(\CA))$.

There are many ways to come up with the required two elements of
the universal exterior algebra in each degree $p$. The natural
construction is a method that works very generally for any finite
group (see below). Alternatively we can use the representatives
implicit in the adhoc computations in Section~2. Thus we lift
$e_x,e_y$ to the universal exterior algebra as elements
\[ \tilde e_x=(1-2X)\extd X, \quad \tilde e_y=(1-2Y)\extd Y\]
where $X,Y$ are some functions as in Section~2. Here
\[\del_x X=1-2X, \quad (1-2X)^2=1, \quad (\extd X)(1-2X)=-(1-2X)\extd
X,\quad [\ddd,X]=(1-2X)R_x\tens\gamma^x\] and similarly for $Y$.
We also need $Z=\delta_{10}+\delta_{01}=X+Y-2XY$ which obeys
\[ (\extd X)(1-2Y)=\extd Z-(1-2X)\extd Y,\quad
[\ddd,Z]=(1-2Z)(R_x\tens\gamma^x+R_y\tens\gamma^y).\] From these
facts it is not hard to compute
\[ \tilde e_y^p=(1-2Y)^{[p]}(\extd Y)^p (-1)^{p(p-1)\over 2},\quad
\pi_p(\tilde e_y^p)=(R_y\tens \gamma^y)^{[p]}\] where $[p]=p$ mod
2, and
\[\tilde e_x \tilde e_y^{p-1}=(1-2X)(\extd X)(1-2Y)^{[p-1]}(\extd
Y)^{p-1}(-1)^{(p-1)(p-2)\over 2},\quad \pi_p(\tilde e_x \tilde
e_y^{p-1})=(R_x\tens\gamma^x)(R_y\tens\gamma^y)^{[p-1]}.\] These
are linearly independent for each $p$ as required. \eproof

A more explicit way to obtain this result, which makes clearer the
quotienting from the universal calculus (and works similarly for
any finite group $G$), is to note that the universal calculus is
itself bicovariant and hence corresponds to some subset, namely
$\CC_{univ}=G-\{e\}$. In our case it means a basic 1-form \bb
e_{x+y}=\sum_g \delta_g\extd \delta_{g+x+y} \ee in addition to
$e_x,e_y$ defined in the same way by (\ref{basic}), but now in
$\Omega_{univ}(\CA)$.  The universal exterior algebra is the free
algebra generated over $\CA$ by these $e_g$ for all $g\in
G-\{e\}$.  Now on any bicovariant calculus (using Hopf algebra
methods) one has a Maurer-Cartan equation, which, for the
universal calculus in our case, comes out as \bb \extd e_x=2
e_x^2+\{ e_x, e_y\}+\{ e_x, e_{x+y}\} -\{ e_y, e_{x+y}\}\ee \bb
\extd e_{x+y}=2 e_{x+y}^2+\{ e_x, e_{x+y}\} +\{ e_y, e_{x+y}\}-\{
e_x, e_y\}\ee and a similar equation for $\extd e_y$. Similarly
for any finite group.

With this description of $\Omega_{univ}(\CA)$ the linear
prolongation exterior algebra mentioned above is just given by
setting to zero all the $e_a$ except  those in our conjugacy
class. In our case we project out $ e_{x+y}=0$ and this yields
the Maurer-Cartan equation for our calculus {\em and} the
additional anticommutation relation, i.e. (\ref{maxrel}) as the
linear prolongation. Likewise the quadratic exterior algebra adds
the additional relation $e_x^2=e_y^2$. Note also that the $\tilde
e_x =e_x+e_{x+y}$ and $\tilde e_y=e_y+e_{x+y}$ used above project
onto the same 1-forms under the quotient as our generators
$e_x,e_y$, but are not so natural from the point of view of the
group structure.

For the products of 1-forms in the universal calculus we note that
$\delta_g(\extd\delta_{g+x})\delta_h=\delta_g\extd(\delta_{g+x}
\delta_h)$,
etc. Hence it is immediate that \bb e_y^p=\sum_g \delta_g\extd
\delta_{g+y}\extd\delta_{g}\extd\delta_{g+y}\extd\delta_g
\cdots,\quad e_xe_y^{p-1}=\sum_g
\delta_g\extd\delta_{g+x}\extd\delta_{g+x+y}\extd\delta_{g+x}\extd
\delta_{g+x+y}\cdots\ee (alternating until the total degree is
$p$). We also need \bb
[\ddd,\delta_g]=(\delta_{g+x}-\delta_g)R_x\tens\gamma^x+(\delta_{g+y}
-\delta_g)R_y\tens\gamma^y.\ee When computing $\pi$ of products of
the $e_x,e_y$ the $\delta_g$ to the front forces which of the four
$\delta$-functions in each $[\ddd,\delta_{\cdot}]$ can
contribute. Let $a,b,c$, etc be chosen from $\{x,y\}$. Then
similarly to the above, we have
\[ e_ae_be_c\cdots=\sum_g\delta_g\extd\delta_{g+a}\extd
\delta_{g+a+b}\extd\delta_{g+a+b+c}\cdots \]
\[\pi(e_a e_b e_c\cdots)=\sum_g\delta_g(R_a\tens\gamma^a)\extd
\delta_{g+a+b}\extd\delta_{g+a+b+c}\cdots\]
\[\quad=(R_a\tens\gamma^a)\sum_g\delta_{g+a}\extd\delta_{g+a+b}
\extd\delta_{g+a+b+c}\cdots=(R_a\tens \gamma^a)\pi(e_b
e_c\cdots)\] after a change of variables. Hence we find for this
description of the universal calculus that \bb \pi(e_a e_b
e_c\cdots)=R_aR_bR_c\cdots\tens \gamma^a\gamma^b\gamma^c\cdots.\ee
In fact we see explicitly that with $\pi(e_x)=R_x\tens\gamma^x$,
$\pi(e_y)=R_y\tens\gamma^y$ and $\pi(e_{x+y})=0$ is an algebra
homomorphism when extended by $\pi(fe_x)=\rho(f)\pi(e_x)$ etc.
Again, this is a general construction for any finite group,
conjugacy class and choice of linearly independent `gamma
matrices'. The map \bb \pi:\Omega_{univ}(\CA)\to
\Omega_\ddd(\CA),\quad \pi(e_a)=\cases{R_a\tens\gamma^a&for\
$a\in \CC$\cr 0&for\ $ a\notin\CC\cup\{e\}$\cr}\ee is an algebra
homomorphism with $\ddd=\sum_a \del_a\tens \gamma^a$. Its kernel
depends on the relations among the gamma-matrices; their only
homogeneous relations being quadratic in our case (and $G$ being
Abelian) is the reason that $\Omega_\ddd(\CA)$ is quadratic.

This completes our algebraic description $\Omega(\CA)$ of the
exterior algebra $\Omega_\ddd(\CA)$ obtained by Connes
construction for our choice of $\ddd$.  The various nonzero
dimensions of the exterior algebra of $\C[\Z_2\times\Z_2]$ are
$1:2:2:2\cdots$ and there is no top form. This means that one
should not expect a Hodge * operator or Poincar\'e duality for
this calculus.

\begin{propos} The quantum de Rham cohomology of this differential
calculus on $\Z_2\times\Z_2$ is
\[ H^0=\C.1,\quad H^1=\C.(e_x-e_y),\quad H^p=\{0\},\quad p\ge 2.\]
\end{propos}
\proof If $f$ is a function and $\extd f=0$ it means $R_x(f)=f$
and $R_y(f)=f$ and hence $f$ is a multiple of the constant
identity function. Hence $H^0$ is spanned by $1$. If a 1-form
(\ref{alpha}) is closed it means (using the Leibniz rule and
$\extd$ as above) that
\[ 0=\extd\alpha=(\del_x\alpha_y-\del_y\alpha_x)e_xe_y
+(\del_x\alpha_x
+\del_y\alpha_y+2\alpha_x+2\alpha_y)e_y^2.\] We write a 2-form
$\alpha_{xy}e_xe_y+\alpha_{yy}e_y^2$ as a vector
$\pp{\alpha_{xy}\cr\alpha_{yy}}$  and $\alpha$ as a vector
$\pp{\alpha_x\cr \alpha_y}$. Then the operator $\extd_1$ which is
$\extd $ on 1-forms is an $8\times 8$ matrix
\[ \extd_1=\pp{\id-R_y&R_x-\id\cr \id+R_x& \id+R_y\cr}\] and its
kernel is easily found to be 4-dimensional. The exact forms in
$\Omega^1$ form a three-dimensional subspace of this kernel (since
$\extd:\Omega^0(\CA)\to\Omega^1(\CA)$ has 1-dimensional kernel
given by constants). Hence $H^1$ is 1-dimensional and easily seen
to be represented by $e_x-e_y$. Also note that the image of
$\extd_1$ is therefore 4-dimensional also. For the general
$\extd_p:\Omega^p\to\Omega^{p+1}$ we note that
\[ \extd e_y^p=\cases{2e_y^{p+1}&for\ $p$\ odd\cr 0&for\
$p$\ even\cr}\]
as one may easily prove by the graded Leibniz rule and induction.
Then
\[ \extd(f e_x e_y^{p-1}+ge_y^p)=(\del_x f+\del_y g+2f)e_y^{p+1}
+g\extd
e_y^{p}+(\del_x g-\del_y f)e_x e_y^{p-1}-fe_x\extd e_y^{p-1}\]
corresponds to the matrix
\[ \extd_p=\pp{(-\id)^{p-1}-R_y& R_x-\id\cr \id+R_x &
(-\id)^{p-1}+R_y\cr}\] which has an order 2 periodicity. In
particular,
\[ \extd_2=\pp{-(\id+R_y)& R_x-\id\cr \id+R_x& -(\id-R_y)}.\] The
transpose of this matrix is easily seen to be conjugate under
 $\pp{0&1\cr -1&0}$ to
 $-\extd_1$. Hence the kernel of $\extd_2$ has the same dimension
as the kernel of $\extd 1$, namely 4. Hence $H^2=\{0\}$. Also the
image of $\extd_2$ is therefore 4-dimensional as is
 the kernel of $\extd_3$ (by periodicity) hence $H^3=\{0\}$. The
 rest vanish by periodicity. \eproof

Let us note as an aside that in the simpler Woronowicz calculus
where we would set $e_x^2=e_y^2=0$, the cohomology by a similar
computation is more easily found to be $H^0=\C.1$,
$H^1=\C.e_x\oplus \C.e_y$ and $H^2=\C.e_xe_y$ which has dimensions
$1:2:1$. This is because in this case the kernel of $\extd$ on
1-forms is 5-dimensional. The exterior algebra in this case also
has the symmetric form with dimensions $1:2:1$ over
$\C[Z_2\times\Z_2]$ but this calculus is not the one coming out
of our Dirac operator using Connes prescription.

\subsection{Gauge theory}

Returning to our above differential calculus, we can also impose
a $*$-structure with $e_x,e_y$ hermitian as for
$\omega_x,\omega_y$ in Section~2. Note that then \bb (\extd
f)^*=e_x (\del_x f)^*+ e_y(\del_y f)^*= e_x\del_x(f^*)+
e_y\del_y(f^*)=-\extd f^*\ee using the definition of $\del$ and
the commutation relations (\ref{dG}). (This is not a property of
the Hilbert space representation). Thus the real cohomology is
$H^0=\R$, etc.

Given a differential calculus one is also free to do `gauge
theory' with connections $\alpha\in \Omega^1(\CA)$. This is
obviously some kind of $U(1)$ gauge theory.  It is worth noting
that from a fully noncommutative geometrical point of
view\cite{BrzMa:gau} it would be better called `U(0)' with $\C$
the enveloping algebra of the zero Lie algebra (or the coordinate
ring of a point). We assume that $\alpha$ is Hermitian, which
means \bb R_x(\alpha_x{}^*)=\alpha_x,\quad
R_y(\alpha_y{}^*)=\alpha_y\ee in terms of its components. Gauge
transformation is by $u\in U(\CA)$ as \bb \alpha^u=u\alpha
u^{-1}+u\extd u^{-1}.\ee In our case a unitary $u$ essentially
means a function on $\Z_2\times\Z_2$ with values in the unit
circle, $u=e^{\imath\phi}$ with $\phi$ real, hence the above is
explicitly \bb\alpha_x\mapsto uR_x(u^{-1})\alpha_x + u\del_x
u^{-1}=e^{-\imath \del_x\phi}\alpha_x+e^{-\imath\del_x\phi}-1,\ee
and similarly for $\alpha_y$. The gauge-covariant curvature
$F(\alpha)=\extd\alpha+\alpha^2$ by a similar calculation to the
above is
\[
F(\alpha)=(2\alpha_x+2\alpha_y+\del_x\alpha_x+\del_y\alpha_y
+\alpha_x
R_x(\alpha_x)+\alpha_y
R_y(\alpha_y))e_x^2+(\del_x\alpha_y-\del_y\alpha_x
+\alpha_xR_x(\alpha_y)
-\alpha_yR_y(\alpha_x))e_xe_y.\] This is just the same result as
in Section~2 except that it is obtained now by working in
$\Omega(\CA)$ and its algebraic relations as above, not by
explicit matrix calculations. Here the two coefficients are
$F_{xx}$ and $F_{xy}$ say, and transform by conjugation of $F$,
which means \bb F_{xx}\mapsto F_{xx},\quad F_{xy}\mapsto
e^{-\imath\del_{x+y}\phi}F_{xy},\ee where $\del_{x+y}=R_xR_y-\id$.

\begin{propos} The moduli space of zero curvature gauge fields
modulo gauge equivalence is a real circle
\[ \lambda^2+\mu^2=2\] modulo $\lambda\mapsto -\lambda$ or
$\mu\mapsto-\mu$. The corresponding gauge fields are
\[ \alpha=(\lambda-1) e_x+(\mu-1)e_y.\]
\end{propos}
\proof  It is easy to see that these are solutions of the
zero-curvature equation, which we leave to the reader. We have to
show that any solution is gauge equivalent to one of these.
First, we change variables to \bb\Phi=\alpha+\theta,\quad
\Phi_a=\alpha_a+1\ee in which case the curvature and gauge
transformation by $u$ have the form \bb
F_{xx}=\Phi_xR_x\Phi_x+\Phi_yR_y\Phi_y-2,\quad F_{xy}
=\Phi_xR_x\Phi_y-\Phi_yR_y\Phi_x ,\quad \Phi_x\mapsto
uR_x(u^{-1})\Phi_x\ee and similarly for $\Phi_y$. The $F=0$
equation clearly becomes \eqn{zeroFPhi}{
\Phi_xR_x\Phi_x+\Phi_yR_y\Phi_y=2,\quad
\Phi_xR_x\Phi_y=\Phi_yR_y\Phi_x.} The first of these implies that
\[ R_x(\Phi_x R_x\Phi_x)=(R_x\Phi_x)\Phi_x,\quad
R_y(\Phi_xR_x\Phi_x)
=R_y(2-\Phi_yR_y\Phi_y)=2-\Phi_yR_y\Phi_y=\Phi_xR_x\Phi_x\] hence
\eqn{lambmu}{ \Phi_x R_x \Phi_x=\lambda^2,\quad \Phi_yR_y \Phi_y
=\mu^2,\quad \lambda^2+\mu^2=2} for some real constants
$\lambda,\mu$. Here the reality property of $\alpha$ translates
as $\Phi_x{}^*=R_x\Phi_x$ etc, and hence $\Phi_x
R_x\Phi_x=|\Phi_x|^2\ge 0$ etc. For the moment we assume that
$\lambda,\mu\ne 0$ and consider the degenerate cases later. Next
we write out the content of the other equation of
(\ref{zeroFPhi}) at the four points of $\Z_2\times\Z_2$,
\eqn{Fxyexpla}{ \Phi_x(0,0)\Phi_y(1,0)=\Phi_y(0,0)\Phi_x(0,1),
\quad \Phi_x(1,0)\Phi_y(0,0) =\Phi_y(1,0)\Phi_x(1,1)}
\eqn{Fxyexplb}{\Phi_x(0,1)\Phi_y(1,1)=\Phi_y(0,1)\Phi_x(0,0),
\quad \Phi_x(1,1)\Phi_y(0,1) =\Phi_y(1,1)\Phi_x(1,0).} In view of
(\ref{lambmu}), most of these equations are redundant and we just
have
\[{ \Phi_x(0,0)\over
\Phi_x(0,1)}={\Phi_y(0,0)\over\Phi_y(1,0)}.\] Let
\[ u(0,0)=1,\quad u(0,1)={\mu\over\Phi_y(0,0)} ,\quad u(1,0)
={\lambda\over\Phi_x(0,0)}, \quad
u(1,1)={\Phi_x(1,1)\mu\over\Phi_y(0,0)\lambda}\] which is unitary
(each component has modulus 1) in view of (\ref{lambmu}). Then
using the above explicit equations and (\ref{lambmu}) one may
verify that
\[ \Phi_x=uR_x u^{-1}\lambda,\quad \Phi_y=u R_y u^{-1}
\mu\] as required. In the special case where $\lambda=0$ we have
$\Phi_x=0$ due to (\ref{lambmu}). We take
\[ u(0,1)=u(1,1)=1,\quad u(0,0)={\Phi_y(0,0)\over\mu},\quad
u(1,0)={\Phi_y(1,0)\over\mu}\] and verify that
$\Phi_y=uR_yu^{-1}\mu$ as required and that $u$ is unitary.
Similarly for $\mu=0$. In these constructions we are free to
choose $\lambda,\mu\ge 0$, for example, but are also free to
choose them in other quadrants of the circle, which means that
the different quadrants are all gauge equivalent to the positive
one. Finally, we consider two gauge fields in our moduli space
for positive $\lambda,\mu$ and $\lambda',\mu'$. If related by a
gauge transformation, we would need ${u(0,0)\over
u(0,1)}\mu=\mu'={u(0,1)\over u(0,0)\mu}$ by looking at
$\Phi_u(0,0)$ and $\Phi_u(0,1)$. These imply that
$\mu'{}^2=\mu^2$ and hence $\mu=\mu'$. Similarly for
$\lambda=\lambda'$ and the degenerate cases. Hence there is
precisely one zero-curvature gauge field up to equivalence for
each parameter pair in the positive quadrant. I.e. the moduli
space is exactly the circle modulo the reflections
$\lambda\mapsto-\lambda,\mu\mapsto -\mu$ (or exactly the quarter
circle with positive values.). \eproof

We see that there is an entire circle of zero curvature gauge
fields with the four quadrants gauge equivalent to each other.
This is the `geometry' of the discrete part of our model. In
particular, the two opposite diameters $\alpha=0$, and
$\alpha=-2\theta$ (or $\Phi=\pm\theta$) are in fact gauge
equivalent. Note also that for the Woronowicz choice with
$e_x^2=e_y^2=0$ the above proof would work in just the same way
since (\ref{Fxyexpla})-(\ref{Fxyexplb}) alone imply that
$|\Phi_x|^2=\lambda^2, |\Phi_y|^2=\mu^2$ as in (\ref{lambmu}) but
without the constraint that the parameters lie on a circle. In
this case the moduli space of zero curvature gauge fields up to
equivalence would be the entire plane modulo the two reflections
(a 1/4 plane), i.e. does not have such a nontrivial topology as
our case.

Finally, coming from the Connes' construction, we have an inner
product particularly on forms. This plays the role of Hodge $*$
and integration against the top form rolled into one (even though
the former does not appear separately). According to Section~2 it
is \bb(f,g)=(fe_x,
ge_x)=(fe_y,ge_y)=(fe_xe_y,ge_xe_y)=(fe_x^2,ge_x^2)=2(f,g)_{l^2}
\ee in terms of the usual $l^2$ inner product on functions $f,g$
(and zero for other combinations of our basic forms). As
explained in Section~2 this defines the gauge field action \bb
\h(F,F)=||F_{xx}||^2+||F_{xy}||^2=|||\Phi_x|^2+|\Phi_y|^2-2||^2
+||R_x(\Phi_x{}^*\Phi_y)-R_y(\Phi_x\Phi_y{}^*)||^2\ee in terms of
the usual $l^2$ norm. Clearly the above $F=0$ solutions form a
circle of minima for this action whose origin is the point
$\alpha=-\theta$ or $\Phi=0$. The points on this circle are not
gauge invariant, being equivalent to their reflections in other
quadrants as well as defining a whole manifold of their further
gauge transforms. According to Section~2 the centre point of the
circle is also an extremum, a local maximum and gauge invariant.
In this way the gauge field action resembles the `Mexican hat'
potential for a Higgs field if we view $\Phi$ as an adjoint higgs
field of some kind rather than as a connection as in our discrete
geometry above.

\section{Particle physics Lagrangians}

In this section, the discrete gauge connections or Higgses $H$
are promoted to genuine fields, i.e. spacetime dependent vectors.
As already in classical quantum mechanics, this promotion is
achieved by tensorizing with functions. Let us denote by $\ff$
the algebra of (smooth, complex valued) functions over 4
dimensional spacetime $M$. Consider the algebra
$\aa_t:=\ff\ot\aa$. The group of unitaries of the tensor algebra
$\aa_t$ is the gauged version of the group of unitaries
$U(\aa)=:\CG$ of the internal algebra $\aa$, i.e. the group of
functions from spacetime into the group $\CG$. Consider the
representation $\rho_t:=\ul\cdot\ot\rho$ of the tensor algebra on
the tensor product $\hh_t:=\sss\ot\hh,$ where $\sss$ is the
Hilbert space of square integrable spinors on which functions act
by multiplication: $ (\ul f\psi)(x):=f(x)\psi(x)$, $ f\in\ff,\
\psi\in\sss$. The spacetime points are labeled $x$ and there
should not be confusions with the discrete label $x\in\zz_2$.
 We denote the Dirac operator on the continuous spacetime  $M$  by
$\ddd_M$ and its chirality operator by $\gamma^5$. The definition
of the tensor product of Dirac operators, \bb\ddd_t:=\ddd_M\ot
1_8+\gamma^5\ot\ddd\ee comes from non-commutative geometry. We now
repeat the above construction for the infinite dimensional
algebra $\aa_t$ with representation $\rho_t$ and Dirac operator
$\ddd_t$. As already stated, for $\aa=\cc,\ \hh=\cc,\ \ddd=0$,
the differential algebra $\Omega_{\ddd_t}(\aa_t)$ is isomorphic
to the de Rham algebra of differential forms $\Omega (M,\cc)$.
For general $\aa$, using the notations of  \cite{sz}, a Hermitian
1-form \bb H_t\in\Omega_{\dd_t}^1(\aa_t),\quad H^*_t=H_t\eee
contains two pieces, a Hermitian Higgs {\it field} $
H\in\Omega^0(M,\Omega_\ddd^1(\aa))$ and a genuine gauge field $
A\in\Omega^1(M,i\rho(\gg))$ with values in $i$ times the Lie
algebra of the group of unitaries,
 \bb \gg:=\left\{ X\in\aa,\ X^*+X=0\right\},\ee represented on
$\hh$. The curvature of $H_t$ \bb C_t:=\de _tH_t+H_t^2\
\in\Omega_{\ddd_t}^2(\aa_t)\ee contains three pieces, \bb
C_t=C+F-\dee\varphi\gamma^5,\ee
 the ordinary, now spacetime dependent curvature
$C=\de H+H^2$, the field strength \bb F:=\de_M
A+\frac{1}{2}[A,A]\quad \in \Omega^2(M,\rho(\gg))\ee and the
covariant derivative of $\varphi$ \bb\dee \varphi=\de_M
\varphi+[A\varphi-\varphi A]
\quad\in\Omega^1(M,\Omega_\ddd^1(\aa)).\ee Note that the covariant
derivative may be applied to $\varphi$ thanks to its homogeneous
transformation law, equation (\ref{hom}).

The definition of the Higgs potential in the infinite dimensional
space $\aa_t$ \bb V_t(H_t):=(C_t,C_t)\ee requires a suitable
regularization of the sum of eigenvalues over the space of
spinors $\sss$. Here we have to suppose spacetime to be compact
and Euclidean. Then, the regularization is achieved by the
Dixmier trace \cite{dix} which allows an explicit computation of
$V_t$. One of the miracles in the Connes-Lott scheme is that
$V_t$ alone reproduces the complete bosonic action of a
Yang-Mills-Higgs model. Indeed, it consists of three pieces,  the
Yang-Mills action, the covariant Klein-Gordon action and an
integrated Higgs potential \bb V_t(A+H)=\int_M\t
\left(F^**F\right)+ \int_M\t
\left(\dee\varphi^**\dee\varphi\right)+ \int_M*V(H).\label{Vt}\ee
As the preliminary Higgs potential $V_0$, the (final) Higgs
potential $V$ is calculated from the finite dimensional triple
$(\aa,\hh,\ddd)$,
   \bb V:=V_0-\t[\alpha C^*\alpha C ]=
\t[(C-\alpha C)^*(C-\alpha C)],\ee where the linear map \bb
\alpha: \Omega_{\ddd}^2(\aa)\longrightarrow
         \rho(\aa)+\pi(\extd \ker\pi_1)\ee is determined by the
two equations \bb \t\lb R^*(C-\alpha C)\rb&=&0\qquad{\rm for\
all}\ R\in\rho(\aa),
\label{a1}\\
 \t\lb K^*\alpha C\rb &=&0\qquad {\rm for\ all}\ K\in\pi(\extd
\ker\pi_1).\label{a2}\ee All remaining traces are over the finite
dimensional Hilbert space $\hh$. We denote the Hodge star by
$*\cdot$. It should not be confused with the involution
$\cdot^*$. Note the `wrong' relative sign of the third term in
equation (\ref{Vt}). The sign is in fact correct for an Euclidean
spacetime.

A similar miracle happens in the fermionic sector, where the
completely covariant action
 $\psi^*(\ddd_t+H_t)\psi$ reproduces the complete fermionic
action of a Yang-Mills-Higgs model. We denote by
 \bb\psi=\psi_R+\psi_L\ \in
\hh_t=\sss\ot\,\left(\hh_R\op\hh_L\right),\qq
\psi_L:=\,\frac{1-\gamma^3}{2}\,\psi,\qq
\psi_R:=\,\frac{1+\gamma^3}{2}\,\psi,\ee
 the multiplets of chiral spinors and by $\psi^*$ the dual of
$\psi$ with respect to the scalar product of the concerned
Hilbert space. We set \bb \ddd_{\CG}=\mm^*\ot\pp{0&1\cr 0&0}+
\mm\ot\pp{0&0\cr 1&0}.\ee $\mm$ will turn out to be the fermionic
mass matrix. Similarly we set \bb H=:\tilde h^*\ot\pp{0&1\cr
0&0}+ \tilde h\ot\pp{0&0\cr 1&0}\in\Omega_\ddd^1(\aa),\ee
\bb\varphi= H+\ddd_{{\CG}}=:\tilde \varphi ^*\ot\pp{0&1\cr 0&0}+
\tilde \varphi \ot\pp{0&0\cr 1&0}\in\Omega_\ddd^1(\aa).\ee Then
\bb\psi^*(\ddd_t+H_t)\psi&=& \int_M*\psi^*(\ddd_M+\gamma(A))\psi
+\int_M*\left(\psi_L^*\tilde h\gamma^5\psi_R +\psi_R^*\tilde
h^*\gamma^5\psi_L\right)\cr\cr
&&+\int_M*\left(\psi_L^*\mm\gamma^5\psi_R
+\psi_R^*\mm^*\gamma^5\psi_L\right)\cr\cr
&=&\int_M*\psi^*(\ddd_M+\gamma(A))\psi
+\int_M*\left(\psi_L^*\tilde \varphi\gamma^5\psi_R +\psi_R^*\tilde
\varphi^*\gamma^5\psi_L\right)\label{diract}\ee containing the
ordinary Dirac action and the Yukawa couplings. Note the unusual
appearance of $\gamma^5$ in the fermionic action (\ref{diract}).
Just as the `wrong' signs in the bosonic action (\ref{Vt}), these
$\gamma^5$ are proper to the Euclidean signature and disappear in
the Minkowski signature. For details see the first reference of
\cite{cl}, example 2, `massless chiral electrodynamics' and
section 6.9, `Wick rotation', of \cite{monsa}.

In our lattice model the junk $\pi(\extd \ker\pi_1)$ is zero and
solving equations (\ref{a1}) and (\ref{a2}) is easy: \bb C-\alpha
C=\rho (\varphi
_xR_x\varphi_y-\varphi_yR_y\varphi_x)\,\omega_x\omega_y,\ee
implying that upon tensorizing with continuous spacetime the
Higgs potential, \bb V=2\{ |\varphi_x(0,0)\varphi_y(1,1)^*-
\varphi _x(1,1)^*\varphi_y(0,0)|^2
+|\varphi_x(0,0)\varphi_y(0,0)^*- \varphi
_x(1,1)^*\varphi_y(1,1)|^2\},\ee loses its precious property of
spontaneous symmetry breaking. This situation is familiar from
the Connes-Lott model of electro-weak forces with one generation
of leptons \cite{cl}, the `minimax example', section 4.6 of
\cite{monsa}.

\section{Discrete diffeomorphisms and spectral action}

Let us summarize Connes' strategy up to this point. He
reformulates Riemannian geometry algebraically in terms of
spectral triples, $(\aa,\hh,\ddd)$. This reformulation is general
enough to never use the commutativity of the algebra $\aa$ of
functions. It is special enough to include generalizations of
differential forms, exterior multiplication and derivative and
the combination of Hodge star and integration needed to define a
Yang-Mills action. On a finite dimensional spectral triple, such a
Yang-Mills action looks generically like a Higgs potential and
breaks the group of unitaries in $\aa$ spontaneously. Tensorizing
the finite dimensional spectral triple with the infinite
dimensional, commutative spectral triple of a Riemannian
manifold, `almost commutative geometry', produces a complete
Yang-Mills-Higgs model. In this setting of almost commutative
geometry, the Higgs scalar is reduced to a pseudo force of the
Yang-Mills force.  This situation is perfectly analogous to
Minkowskian geometry (special relativity) reducing the magnetic
force to a pseudo force of the electric force.

With his fluctuating metric, Connes goes one step further
\cite{co}. His algebraic reformulation of Riemannian geometry of
course contains a generalization of the Riemannian metric, the
Dirac operator $\ddd$. This generalization is special enough to
allow for an algebraic reformulation of general relativity in
terms of the commutative spectral triple of a Riemannian
manifold. The kinematical part of this algebraic reconstruction
is the fluctuating metric, the dynamical part is the spectral
action \cite{cc}. Repeating this algebraic construction for almost
commutative spectral triples produces in addition to general
relativity some very special Yang-Mills-Higgs models. In this
almost commutative setting therefore these very special
Yang-Mills-Higgs forces are reduced to pseudo forces of gravity.
The electromagnetic, weak and strong forces are among these very
special Yang-Mills-Higgs forces.

The central tool to construct the fluctuating metric is the lift
of the group of automorphisms and unitaries of $\aa$ to the
Hilbert space $\hh$. For the commutative triple of a Riemannian
manifold, the automorphisms are the diffeomorphisms of the
manifold interpreted as general coordinate transformations and
their image under the lift are the local spin transformations.
The unitaries are gauged $U(1)$ transformations. In presence of
the real structure, they are all lifted to the identity.  Let us
compute the lift in our lattice example. The automorphism group
of our algebra $\aa=\cc[\zz_2\times\zz_2]$ is \bb {\rm
Aut}(\aa)=S_4\owns P,\ee the group of permutations of the four
points. It is the discrete version of the diffeomorphism group.
We disregard complex conjugation, that is we do not consider
$\aa$ to be real. The group of unitaries \bb U(\aa)=U(1)^4\owns
u(x,y)\ee is the discrete version of Maxwell's gauge group.
Simultaneously it plays the role of the gauged Lorentz group. We
need to map both groups to the group of automorphisms lifted to
the Hilbert space $\hh$, \bb {\rm
Aut}_\hh(\aa):=\{U:\hh\rightarrow\hh,\ UU^*=U^*U=1,\ [U,\gamma
_3]=0,\forall f\in \aa\ U\rho(f)U^{-1}=\rho(\tilde f),\
\exists\tilde f\in\aa\}.\!\!\!\!\!\!\!\!\cr \ee As mentioned our
example does not satisfy Connes' first order condition. Anyway we
would have a hard time to choose the sign of the square of the
real structure since this square is plus one in dimension zero,
minus one in dimension two. Therefore we do not introduce a real
structure in the definition of the lifted automorphisms. Every
lifted automorphism $U$ projects down to an automorphism $P=p(U)$
with $P(f)=\tilde f$. In our example we have \bb {\rm
Aut}_\hh(\aa)=S_4\ltimes \left( U(1)_L^4\times
U(1)_R^4\right)\owns(P,u_L(x,y), u_R(x,y)). \ee Let us denote the
lifting homomorphism by $(L,\ell):{\rm Aut}(\aa)\ltimes
U(\aa)\longrightarrow{\rm Aut}_\hh(\aa).$ It must satisfy
$(p\circ (L,\ell))(P,u)=P.$ Let us start with the automorphisms
alone, $L(P)=(\beta (P),u_L(P), u_R(P))$. The most general
solution is $\beta (P)=P$, $u_L(P) =\sigma_L(P)\,1_4$, $u_R(P)
=\sigma_R(P)\,1_4$ where the two functions $\sigma_{L,R}:
S_4\rightarrow \zz_2$ are either identically one or the signature
of the permutation, four possibilities. We have written an $1_4$
to indicate that the unitaries are rigid, i.e. independent of $x$
and $y$. As unitary $8\times 8$ matrices the four possible lifts
take the form: \bb L(P)= P\ot \left[{\textstyle\frac{\sigma_L+
\sigma_R}{2}}1_2\,+\,{\textstyle\frac{\sigma_L-
\sigma_R}{2}}\gamma^3\right].\ee They only induce trivial
fluctuations of the metric, \bb L(P)\ddd L(P)^{-1}=\pm(\tilde
\pa_x\ot\gamma^x+ \tilde \pa_y\ot\gamma^y),\qq \tilde
\pa_\cdot=P\pa_\cdot P^{-1}.\ee This is in sharp contrast to the
continuous case where the lifted diffeomorphisms induce the
general curved metric starting from the flat one. Fortunately
upon tensorizing with a continuous spacetime we obtain a general
internal Dirac operator that acquires the status of the fermionic
mass matrix. In the almost commutative setting, we will also see
the lift of the unitaries of our internal algebra
$\aa=\cc[\zz_2\times\zz_2]$.

The automorphisms of $\aa_t=\ff\ot\aa$ close to the identity are
diffeomorphisms of spacetime $\phi\in {\rm Diff}(M) $. The group
of unitaries $U(\aa_t)$ is the gauged $^MU(1)^4$ whose elements
are functions from $M$ to $U(1)^4$ that we denoted by
$u=(u(0,0),u(1,0),u(0,1),u(1,1))$ as before. The group of
automorphisms lifted to the Hilbert space has as component
connected to the identity \bb{\rm Aut}_{\hh_t}(\aa_t)={\rm
Diff}(M) \ltimes \,^M\left(Spin(4)\times U(1)_L^4\times
U(1)_R^4\right)\owns(\phi,u_L, u_R).\ee The lift $L(\phi)$ is
described explicitly in \cite{lift} and locally it induces the
general curved Dirac operator on $M$ by fluctuating the flat one,
\bb L(\phi)\ddd_{\rm flat}L(\phi)^{-1}= i\hbox{$
e^{-1\,\mu}$}_a\gamma^a\left[ \pa /\pa {x^{
\mu}}+{\textstyle\frac{1}{4}}\omega _{bc\mu}\gamma^b\gamma^c
\right]= \ddd_M,\ee with tetrad coefficients ${e^a}_\mu$ and
their torsionless spin connection 1-form $\omega_{bc\mu}\de
x^\mu$. Let us concentrate on lifting the unitaries:
$\ell(u)=\rho_t(u)$, i.e. $u_L=u_R$ meets all requirements:
$\ell$ is a group homomorphism, and for every unitary $u$ in
$U(\aa_t)$, $\ell(u)$ is a unitary operator on $\hh_t$, $\ell(u)$
commutes with $\gamma^5\ot\gamma^3$ and $p\circ
\ell(u)=1_{\aa_t}.$ We are ready to fluctuate the metric again,
\bb &&\ell(u)\ddd_t\ell(u)^{-1}=:\ddd_{t\ \rm fluct}\cr =
&&i\hbox{$ e^{-1\,\mu}$}_a\gamma^a\left[ \pa /\pa {x^{ \mu}}\ot
1_8+{\textstyle\frac{1}{4}}\omega _{bc\mu}\gamma^b\gamma^c \ot
1_8-1_4\ot i \rho(A_\mu) \right]+
\gamma^5\ot\left[H+\ddd\right]\cr =&&i\hbox{$
e^{-1\,\mu}$}_a\gamma^a\left[ \pa /\pa {x^{ \mu}}\ot
1_8+{\textstyle\frac{1}{4}}\omega _{bc\mu}\gamma^b\gamma^c \ot
1_8-1_4\ot i\rho(A_\mu) \right]+ \gamma^5\ot\varphi,\ee with the
Yang-Mills connection 1-form $i A_\mu\de x^\mu=u\de u^{-1}$. As
in the Connes-Lott scheme, the Higgs scalar appears as a
connection 1-form with respect to the internal spectral triple,
$H=\pi (u\extd u^{-1})/i=:\varphi-\ddd$.  As before we expand
$\varphi=:\rho(\varphi_x)\omega_x+ \rho(\varphi_y)\omega_y$ with
four, now spacetime dependent complex coefficients,
$\varphi_x(0,0)=\varphi_x(1,0)^*,$
$\varphi_x(1,1)=\varphi_x(0,1)^*,$
$\varphi_y(0,0)=\varphi_y(0,1)^*,$
$\varphi_y(1,1)=\varphi_y(1,0)^*.$ The kinematics is defined by a
metric encoded in $\ddd_M$ or its tetrad coefficients, by  a
Yang-Mills potential, i.e. a 1-form $A$ with values in $i$ times
the Lie algebra of $U(\aa)$ and by four complex Higgs scalars.

In general relativity, the dynamics of the metric is essentially
fixed by a diffeomorphism invariant action functional. In the
setting of spectral triples, there is a natural automorphism
invariant action functional, the trace of the fluctuated Dirac
operator, i.e. of the Dirac operator, that is minimally coupled
to the metric, to the Yang-Mills potential and to the Higgs
scalars. Since the Dirac operator is self adjoint and anticommutes
with the chirality $\gamma^5\ot\gamma^3$, its spectrum is even
and it is enough to compute the trace of its square. Being
divergent, this trace is regularized by a function
$f:\rr_+\rightarrow\rr_+$ of sufficiently fast decrease and the
celebrated spectral action of Chamseddine \& Connes \cite{cc}
reads: \bb S[g,A,\Phi]=\t f(\ddd_{t\ \rm fluct}^2/\Lambda). \ee
For convenience we have put in a scale factor $\Lambda$ carrying
the dimension of the eigenvalues of the Dirac operator, say GeV.
Asymptotically for large $\Lambda$, the spectral action
reproduces the Einstein-Hilbert action and a complete
Yang-Mills-Higgs action. In this limit the regularizing function
$f$ is universal in the sense that the spectral action only
depends on its first three `moments', $f_0:=\int_0^\infty
tf(t)\de t$, $f_2:=\int_0^\infty f(t)\de t$ and $f_4=f(0)$. In
particular its Higgs potential is: \bb V=\lambda\,\t
_8(\varphi^*\varphi\varphi^*\varphi)-\mu^2/2\,\, \t
_8(\varphi^*\varphi),\qq \lambda=\pi /f_4,\qq
\mu^2/2=(f_2/f_4)\Lambda^2  .\ee A straight forward calculation
gives, \bb V&=\ 2\lambda\,\{&\ \ [|\varphi_x(0,0)|^2
+|\varphi_y(0,0)|^2]^2 +[|\varphi_x(0,0)|^2 +|\varphi_y(1,1)|^2]^2
\cr &&+[|\varphi_x(1,1)|^2 +|\varphi_y(0,0)|^2]^2
+[|\varphi_x(1,1)|^2 +|\varphi_y(1,1)|^2]^2\cr &&
+2|\varphi_x(0,0)\varphi_y(1,1)^*- \varphi
_x(1,1)^*\varphi_y(0,0)|^2 \cr &&
+2|\varphi_x(0,0)\varphi_y(0,0)^*-
\varphi_x(1,1)^*\varphi_y(1,1)|^2\}\cr &-\mu^2\{ &\ \
\varphi_x(0,0)^*\varphi_x(0,0) +\varphi_x(1,1)^*\varphi_x(1,1)\cr
&&+\varphi_y(0,0)^*\varphi_y(0,0)
+\varphi_y(1,1)^*\varphi_y(1,1)\} .\ee  As its brother from
section 2, equation (\ref{brother}), this potential has
continuously degenerate minima,
$\varphi_x(0,0)=\varphi_x(1,1)=\mu/(2\sqrt\lambda) \sin\beta $,
$\varphi_y(0,0)=\varphi_y(1,1)=\mu/(2\sqrt\lambda)\cos\beta$. All
minima break the gauged $^MU(1)^4$ spontaneously down to a
single, rigid $U(1)$, except when $\beta$ is an integer multiple
of $\pi/2 $. Then the little group is $U(1)^2$.

\section{Concluding remarks}

We conclude the paper with a brief outline, using again our
quantum group methods, of what happens for other lattices \bb
G=(\Z_m)^n.\ee Clearly one might turn to these for better
approximations of $n$-dimensional tori.

We take $\CA=\C[(\Z_m)^n]$ of course and the usual $n$-dimensional
$\gamma$-matrices $\gamma^i$, $i=1,\cdots, n$. The calculus has
the allowed directions which are the standard basis vectors
$\CC=\{\vec{x}_i|\ i=1,\cdots n\}$ of the lattice,  where
$\vec{x}_i=(0,\cdots,0,1,0\cdots 0)$ denotes the element of
$(\Z_m)^n$ with  1 in the $i$'th place. Thus $\Omega^1(\CA)$ is
spanned by $\{e_i|\ i=1,\cdots,n\}$, where $e_i=: e_{\vec{x_i}}$
is a shorthand. Likewise $\del_i=R_{e_{\vec{x_i}}}-\id$ is the
lattice differential in the $i$'th direction  in $(\Z_m)^n$. This
description is necessarily isomorphic to the 1-forms in Connes
construction for $\ddd=\sum_i \del_i\tens\gamma^i$.

For the higher forms we first compute the linear prolongation of
$\Omega^1(\CA)$. Whatever  the Connes $\Omega_\ddd(\CA)$ is, it
must be a quotient of this. Using the method of  Section~3.1 we
start with the universal exterior algebra with generators
$\{e_{\vec g}|\ \vec{g}\in(\Z_m)^n,\ \vec{g}\ne 0\}$. The linear
prolongation consists of  setting to zero all except the
$\{e_i\}$. However, the Maurer-Cartan equations in the  universal
exterior algebra are \bb \extd
e_{\vec{g}}=\{\theta_{univ},e_{\vec{g}}\}-\sum_{\vec{b}
+\vec{c}=\vec{g},\
\vec{b}, \vec{c}\ne 0}e_{\vec{b}}e_{\vec{c}},\quad
\theta_{univ}=\sum_{\vec{g}\ne 0}e_{\vec{g}}.\ee This is a special
case of the Maurer-Cartan equations for any Hopf algebra and in
any case  easily verified from the standard form of the
$e_{\vec{g}}$ in terms of $\delta$-functions  on $G$. Projecting
out all but the $\{e_{i}\}$ gives \bb \extd
e_i=\{\theta,e_i\},\quad \theta=\sum_i e_i\ee and \eqn{mc2}{
0=\sum_{\vec{x}_i+\vec{x}_j=\vec{g}}e_i e_j,\quad \forall
\vec{g}\in(\Z_m)^n, \ \vec{g}\ne 0.}

In all these equations, addition of vectors is mod $m$. If $m>2$
the equation (\ref{mc2})  has two non-empty cases. When
$\vec{g}=2\vec{x}_i$ for some $i$, we have the equation \bb
e_i^2=0\ee and when $\vec{g}=\vec{x_i}+\vec{x_j}$ for some $i\ne
j$ we have \bb \{e_i,e_j\}=0.\ee Hence in this case the linear
prolongation already coincides with the Woronowicz exterior
algebra, which in turn is the `trivial' one similar to that of
$\R^n$. The Connes exterior algebra cannot have stronger
relations than this and hence this is also $\Omega_\ddd(\CA)$ in
this case.  In particular, $e_i^2=0$ eliminates all of the
interesting features of our model such as the Higgs potential and
spontaneous symmetry breaking. The model in effect resembles
more like flat space.

On the other hand $m=2$ is precisely the case where
$2\vec{x_i}=0$ and is therefore not  one of the possibilities for
$\vec{g}$ in (\ref{mc2}). Thus in this case the linear
prolongation has only the relation $\{e_i,e_j\}=0$ for $i\ne j$,
in particular $e_i^2\ne 0$ as for our $\Z_2\times\Z_2$ case. We
also have $\ddd$ Hermitian and the  same properties for the
$\del_i$ as in the $n=2$ case. In particular, we have the  same
features of the Higgs potential etc. Finally, since $(R_i)^2=\id$
as before, we have $\pi(e_i^2) =(R_i\tens\gamma^i)^2=1$ and
similar features for the higher forms. In summary, our $\Z_2\times
\Z_2$ model is  typical of the general $(\Z_2)^n$ for $n\ge 2$.

Finally, we remark that the methods in this paper do apply to
other finite groups just  as well. For example, they could also
be applied to a nonAbelian group or `curved  lattice' as in
\cite{Ma:rieq}\cite{MaRai:ele}. The first of these papers also
proposes a general choice of $\gamma$-matrices (namely built from
an irreducible representation of the finite group) and explicitly
proposes a Dirac operator for the permutation group $S_3$ in this
way. Development of that model along similar lines to that here
could  be an interesting topic for further work.

 \end{document}